\documentclass[letterpaper, 10 pt, conference]{ieeeconf}  

\IEEEoverridecommandlockouts                              

\usepackage{graphicx} 

\title{\LARGE \bf
Safe Q-learning for continuous-time linear systems }
\author{Soutrik Bandyopadhyay and Shubhendu Bhasin
\thanks{The authors are with the Department of Electrical Engineering, Indian Institute of Technology Delhi, New Delhi, India.
        {\tt\small \{Soutrik.Bandyopadhyay,sbhasin\}@ee.iitd.ac.in}}
}

\usepackage{amsmath,amsfonts,amssymb}
\usepackage{hyperref}

\newtheorem{assm}{Assumption}
\newtheorem{thm}{Theorem}

\newtheorem{defn}{Definition}

\newtheorem{rem}{Remark}

\newcommand{\quadr}[2]{#2^{\Tr} #1 #2}
\newcommand{\norm}[1]{\|#1\|}

\renewcommand{\Re}{\mathbb{R}}
\newcommand{\Tr}{\top}

\newcommand{\proj}{\text{proj}}
\newcommand{\trace}{\text{tr}}
\newcommand{\Bs}{B_{s}}
\newcommand{\gradBs}{\nabla_{x} B_{s}}
\newcommand{\gradBsT}{\nabla_{x} B_{s}^{\Tr}}

\newcommand{\uSafe}{u^{*}_{\text{safe}}}
\newcommand{\uHat}{\hat{u}_{\text{safe}}}

\newcommand{\WaHat}{\hat{W}_{a}} \newcommand{\WaTilde}{\tilde{W}_{a}}
\newcommand{\WaHatDot}{\dot{\hat{W}}_{a}}

\newcommand{\WcHat}{\hat{W}_{c}} \newcommand{\WcTilde}{\tilde{W}_{c}}
\newcommand{\WcHatDot}{\dot{\hat{W}}_{c}}
\newcommand{\WcTildeDot}{\dot{\tilde{W}}_{c}}

\newcommand{\C}{\mathcal{S}}
\newcommand{\IntC}{\text{Int}(\mathcal{S})}
\newcommand{\dC}{\partial\mathcal{S}}

\newcommand{\fall}{\;\;\forall\;}

\newcommand{\VStar}{V^{*}_{s}}

\newcommand{\gradVStar}{\nabla_{x} \VStar}

\newcommand{\Qxx}{Q_{11}}
\newcommand{\Qxu}{Q_{12}}
\newcommand{\Qux}{Q_{21}}
\newcommand{\Quu}{Q_{22}}

\newcommand{\QuxHat}{\hat{Q}_{21}}
\newcommand{\QuuHat}{\hat{Q}_{22}}

\newcommand{\QuxTilde}{\tilde{Q}_{21}}

\newcommand{\lamHat}{k_{sb}}

\newcommand{\eigMin}{\lambda_{\text{min}}}
\newcommand{\eigMax}{\lambda_{\text{max}}}

\newcommand{\upperB}{\overline{\mathcal{B}}}

\DeclareMathOperator*{\argmin}{arg\,min}

\begin{document}

\maketitle
\thispagestyle{empty}
\pagestyle{empty}

\begin{abstract}
Q-learning is a promising method for solving optimal control problems for uncertain systems without the explicit need for system identification. However, approaches for continuous-time Q-learning have limited provable safety guarantees, which restrict their applicability to real-time safety-critical systems. This paper proposes a safe Q-learning algorithm for partially unknown linear time-invariant systems to solve the linear quadratic regulator problem with user-defined state constraints. We frame the safe Q-learning problem as a constrained optimal control problem using reciprocal control barrier functions and show that such an extension provides a safety-assured control policy. To the best of our knowledge, Q-learning for continuous-time systems with state constraints has not yet been reported in the literature.
\end{abstract}


\section{Introduction}
Reinforcement learning(RL) has a strong inter-relationship with the theory of
adaptive optimal control \cite{sutton1992reinforcement}. In particular, RL
algorithms have seen reasonable success in solving continuous-time optimal
control problems for systems with uncertain/unknown dynamics (see
\cite{bhasin2013Automatica,vamvoudakis2010Automatica,lewis2012CSM,kamalapurkar2016Automatica,kamalapurkar2016model}
and references therein for some examples). Stemming from the theory of dynamic programming for continuous-time systems, such approaches typically solve the Hamilton-Jacobi-Bellman(HJB) equations \cite{lewis2012optimal} under uncertain
system dynamics by observing system trajectories. However, unlike its
discrete-time counterpart, the Bellman equation, the HJB equation
requires accurate knowledge of the system dynamics. Thus, solving HJB equations
for continuous-time uncertain systems
involve some degree of system identification to identify the unknown/uncertain
system dynamics.

One promising approach to solving optimal control problems without exact
knowledge of the
system dynamics is Q-learning \cite{watkins1992ML}. Inspired by algorithms for
discrete-time Q-learning
\cite{hasselt2010NIPS,tsitsiklis1994ML,abounadi2002SIAM,bhatnagar2009Automatica,bradtke1992NIPS},
significant research effort is directed towards extending Q-learning to
continuous-time optimal control problems
\cite{jha2014ADPRL,vamvoudakis2017SysConLet,mehta2009CDC,kim2021JMLR,kim2020L4DC,lee2012Automatica,palanisamy2014continuous}.
Such approaches have shown promising results in learning optimal control
policies without needing to know the exact system dynamics.
However, applying such algorithms to real-time safety critical systems is still
an open challenge due to lack of safety guarantees.

Formally, the notion of
safety of dynamical systems is the certification of forward
invariance\cite{blanchini1999Automatica} of state and actuation constraint sets.
Under this definition of safety, the safe RL problem is the
mathematical construct to solve optimal control problems under user-defined
state and actuation constraints. In the literature, some common approaches to
ensure safety, include
model predictive control (MPC)
\cite{zanon2020TAC,wabersich2021TAC,li2018ACC},
reachability analysis
\cite{fisac2018general,fisac2019ICRA}
and control barrier functions
\cite{ames2016TAC,ames2019ECC}
to name a few. Control barrier functions have
gained popularity recently because they provide a Lyapunov-like analysis to study a system's safety without the need to compute the system trajectories.

In the literature, the state and input-constrained linear quadratic
regulation (LQR) problem has been extensively studied
\cite{scokaert1998TAC,johansen2000CDC,dean2019ACC,bemporad2002Automatica}.
Further, approaches that combine
adaptive and optimal control theory to solve the LQR problem for uncertain systems
have also been reported
\cite{jiang2012Automatica,jha2017IFAC,jha2019TAC,modares2014TAC}. However, solving constrained LQR problems under uncertain system
dynamics is still an open challenge.

A particular class of solutions for the constrained adaptive optimal control
problem can be found in
\cite{cohen2020CDC,cohen2023Automatica,marvi2021IJRNC,mahmud2021ACC} where
control policies are learned via a constrained approximate dynamic programming
approach. However, all these approaches typically require an online system
identification to identify uncertain system dynamics. This requirement for
system identification comes at a price of increased computation complexity for these approaches.
As discussed before, some continuous-time Q-learning approaches have shown the
ability to learn optimal control policies without needing this online system
identification and thus, they are computationally cheaper.

In the context of continuous-time Q learning, the authors of
\cite{zhang2020QLearningMPC} have applied
MPC to Q-learning in order to incorporate actuation constraints. The authors of
\cite{modares2014Automatica} discuss an integral reinforcement learning
technique with input constraints.
Continuous-time Q-learning has been used for kino-dynamic motion planning in
\cite{kontoudis2019TNNLS}.

To the best of our knowledge, continuous-time Q-learning with state constraints
has not been reported in the literature. In this paper, we propose a safe
Q-learning algorithm to handle user-defined state constraints using
reciprocal control barrier functions \cite{ames2016TAC}.

\subsection{Contributions}
This work extends the
continuous-time Q-learning framework to incorporate state constraints. The
distinct advantage of the proposed method over constrained approximate dynamic
programming approaches is that it does not require
an explicit system identification step while safely learning the optimal control
policy.

We first formulate the safe Q-learning problem as the constrained optimization
of the Q function with respect to the control policy, subject to the constraint
on the time derivative of a reciprocal control barrier function. We subsequently formulate the Lagrangian of the optimization problem and use
an analytical solution to compute a constrained optimal control policy.
We show that the proposed method bridges the gap between constrained adaptive
optimal control and the ad-hoc method of safeguarding controllers \cite{cohen2023Automatica}.
We then extend the integral reinforcement learning technique to safely learn the
optimal control policy online and show that the proposed method satisfies the
user-defined state constraints.

\subsection{Mathematical notations used}
In this paper, we use $\succcurlyeq$ and $\succ$ to denote square matrices' semi-definite and
definite ordering, respectively. For a function
$(\cdot) : \Re^{n} \rightarrow \Re$, $\nabla_{x}(\cdot)$ denotes
$\frac{\partial (\cdot)}{\partial x}$.
We use
$\mathbb{N}_{n}$ to denote the set of all natural numbers up to and including
$n$.
We use $\eigMax(\cdot)$ and $\eigMin(\cdot)$ to refer to the maximum and minimum
eigenvalues of a square matrix, respectively.
We use $\norm{\cdot}$ to denote the 2-norm for
vectors and the corresponding induced norm for matrices. Additionally,
$\trace(\cdot)$ denotes the trace of a square matrix.
\section{Problem Formulation and Preliminaries}
Consider the linear time-invariant system
\begin{equation}
  \label{eq:plant}
  \dot{x}(t) = A x(t) + B u(t) , \;\; x(0) = x_{0},
\end{equation}
where $x(t) \in \Re^{n}$ is the system state,
$u(t) \in \Re^{m}$ is the control input,
$A \in \Re^{n \times n}$ is the uncertain system matrix, and
$B \in \Re^{n \times m}$ is the input matrix. We assume that
pair $(A, B)$ is controllable and $B$ is full rank and known.
For the system in \eqref{eq:plant}, we seek to solve the infinite-horizon linear
quadratic regulation (LQR) problem by minimizing the
cost functional
\begin{equation}
  \label{eq:costFunction}
  J(x(0),u) \triangleq \frac{1}{2}\int_{0}^{\infty} \quadr{M}{x(\tau)}
  + \quadr{R}{u(\tau)} d\tau,
\end{equation}
with respect to the control policy $u$, where $M \in \Re^{n\times n}$ and
$R \in \Re^{m \times m}$ are state and input weighing matrices respectively. The
matrix $M$ is positive semi-definite and the pair $(\sqrt{M}, A)$ is
detectable. The control weighing matrix $R$ is positive definite. Additionally, we impose the
following safety constraints on the state trajectory of the system
\begin{equation}
  x(t) \in \C \fall t \in \Re_{\ge 0},
\end{equation}
where the set $\C$ is a user-defined compact set containing the origin. In other
words, the control policy must ensure the forward invariance of the set $\C$
\cite{blanchini1999Automatica}. For the rest of the paper, we suppress the time
dependence of the signals $x(\cdot)$ and $u(\cdot)$ for notational brevity.

\subsection{Unconstrained optimal control}
For the system in \eqref{eq:plant} and the cost functional in
\eqref{eq:costFunction}, the Hamiltonian \cite{lewis2012optimal} is defined as
\begin{equation}
\label{eq:hamiltonian}
\begin{aligned}
H(x,u,&\gradVStar(x)) \triangleq
\frac{1}{2}(\quadr{M}{x}
+\quadr{R}{u}) \\&
+ \gradVStar(x)^{\Tr}(Ax + Bu), \fall (x,u) \in \Re^{n}\times \Re^{m},
\end{aligned}
\end{equation}
where $\VStar(x) : \Re^{n} \rightarrow \Re$ is the optimal value function
defined as
\begin{equation}
\label{eq:valueFunction}
\begin{aligned}
\VStar(x(t)) \triangleq
\min_{u(\tau) \forall \tau \in \Re_{\ge 0}}\int_{t}^{\infty}
\frac{1}{2}(\quadr{M}{x(\tau)}
  \\
  + \quadr{R}{u(\tau)})
d \tau.
\end{aligned}
\end{equation}
The optimal control law for the unconstrained system is obtained by
minimizing the Hamiltonian with respect to (w.r.t.)
the control action for each state $x \in \Re^{n}$, i.e.,
\begin{equation}
  \label{eq:uFromHamiltonian}
  \begin{aligned}
  u^{*}(x)
  &= \argmin_{u} H(x,u,\gradVStar(x)) \\&
  = -R^{-1} B^{\Tr} \gradVStar(x).
  \end{aligned}
\end{equation}
For the case of LTI systems, under quadratic integral cost functionals, it is well known that the value function is a quadratic function of the state \cite{lewis2012optimal}, i.e.,
\begin{equation}
  \label{eq:VStar}
\VStar(x) = \frac{1}{2}\quadr{P}{x} ,
\end{equation}
where $P \in \Re^{n\times n}$ is a unique positive definite symmetric matrix obtained
by solving the algebraic Riccati equation (ARE)
\begin{equation}
\label{eq:are}
A^{\Tr}P + PA - PBR^{-1}B^{\Tr}P + M = 0,
\end{equation}
and the optimal control from \eqref{eq:uFromHamiltonian} takes the form
\begin{equation}
\label{eq:unsafeOptimalControl}
u^{*}(x) = - R^{-1} B^{\Tr} Px.
\end{equation}

The solution to the ARE in \eqref{eq:are} and
the corresponding optimal control law in \eqref{eq:unsafeOptimalControl} require
complete knowledge of the system matrices $A$ and $B$. To solve optimal control problems
for systems with uncertain/unknown
dynamics, continuous-time approximate dynamic programming (ADP) approaches have
been proposed in the literature
\cite{jiang2012Automatica,bhasin2013Automatica,vamvoudakis2010Automatica}.

\subsection{Continuous-time Q-learning}
A notable approach for solving the ARE in a model-free setting is to
define the so-called ``$Q$-function'' inspired by the field of reinforcement
learning in discrete-time setting \cite{bradtke1992NIPS,watkins1989learning,watkins1992ML}. The advantage of
this method is that the optimal control policy can be learned online from the state
observations without needing to know the system dynamics.



In the present work, we define the function
$Q: \Re^{n}\times \Re^{m} \rightarrow \Re$ as
\cite{vamvoudakis2017SysConLet,vamvoudakis2015Automatica}
\begin{equation}
  Q(x,u) \triangleq \VStar(x) + H(x,u, \gradVStar).
\end{equation}
Substituting the value function from \eqref{eq:valueFunction} and the
Hamiltonian yields
\begin{equation}
  \label{eq:QStarDefinition}
  Q(x,u) = \frac{1}{2}
  X^{\Tr}
\underbrace{
\begin{bmatrix}
  \Qxx & \Qxu \\
  \Qux & \Quu
\end{bmatrix}
}_{\triangleq \; \overline{Q}}
X,
\end{equation}
where $X \triangleq [x^{\Tr} \; u^{\Tr}]^{\Tr}$,
$\Qxx \triangleq PA + A^{\Tr}P + P + M $,
$\Qxu = \Qux^{\Tr} \triangleq PB$,
and $\Quu \triangleq R$ are matrices of
appropriate dimensions (cf. \cite{vamvoudakis2017SysConLet}). Based on this definition of the $Q$ function, the
optimal control law $u^{*}:\Re^{n} \rightarrow \Re^{m}$, can be written as
\begin{equation}
\label{eq:qbasedcontrol}
u^{*}(x) = \argmin_{u}Q(x,u) = -\Quu^{-1}\Qux x.
\end{equation}
The expression in \eqref{eq:qbasedcontrol} offers a possible way to approximate
the optimal control in a model-free way using the estimates of $\Quu$ and $\Qux$
\cite{vamvoudakis2017SysConLet}.
In this paper, we extend the above formulation to incorporate user-defined
safety constraints by utilizing Lyapunov-like control barrier functions.

\subsection{Control barrier functions}
A versatile approach to ensure the safety of dynamical systems is via control
barrier functions, which are Lyapunov-like functions used to provide safety
certificates to control policies \cite{ames2016TAC,ames2019ECC,taylor2020ACC}.
Specifically, let there exist a continuously differentiable function
$h(x): \Re^{n} \rightarrow \Re $, such that
\begin{align}
\C &= \{x \in \Re^{n} : h(x) \ge 0 \},\\
\IntC &= \{x \in \Re^{n} : h(x) > 0 \},\\
\dC &= \{x \in \Re^{n} : h(x) = 0 \}.
\end{align}
where $\IntC$ and $\dC$ are non-empty sets defined as the interior and
the boundary of the compact set $\C$, respectively. The function $h(x)$ is often
referred to as the ``zeroing'' control barrier function (ZCBF).
In this paper, we consider another type of control barrier function, namely - reciprocal control barrier function (RCBF)
\cite{ames2016TAC} due to its similarities with Lyapunov functions.
The RCBF is defined as
\begin{defn}[Reciprocal control barrier function \cite{ames2016TAC}]
A continuously differentiable function
$\Bs(x): \IntC \rightarrow \Re$ is said to be a RCBF for the system in
\eqref{eq:plant} if there exist class
$\mathcal{K}$ functions
$\alpha_{1},\alpha_{2},\alpha_{3}$ such that
\begin{align}
  &\frac{1}{\alpha_{1}(h(x))}
  \le \Bs(x) \le
    \frac{1}{\alpha_{2}(h(x))},\label{eq:boundsOfBarrier} \\&
  \inf_{u}[\gradBs(x)^{\Tr} (Ax + Bu)] \le \alpha_{3}(h(x)), \fall x \in \C.
\end{align}
\end{defn}

Provided a valid RCBF $\Bs(x)$ exists, a control policy
$u(x): \IntC \rightarrow \Re^{m}$ satisfying
\begin{equation}
  \label{eq:safetyConstraint}
  \gradBs(x)^{\Tr}[Ax + Bu(x)] \le \gamma(1/\Bs(x)) \fall x \in \IntC,
\end{equation}
for some class $\mathcal{K}$ function $\gamma(\cdot)$; renders the set $\C$ forward invariant for the system \eqref{eq:plant} \cite{ames2016TAC}.
We assume that for the given safety constraint set $\C$, there exists a valid
Lyapunov-like RCBF for the system in \eqref{eq:plant}. We now use RCBF to ensure
safe online training of the continuous-time Q-learning algorithm.







\section{Safe Q-Learning}
\label{sec:sql}
We now detail the main contribution of the present work.
The objective of the proposed safe Q-learning algorithm is to modify the optimal control policy of the unconstrained problem in a minimally-invasive fashion to ensure
safety. Thus, we qualify the optimization problem in \eqref{eq:qbasedcontrol} by
the safety constraint \eqref{eq:safetyConstraint} and formulate the safe Q-learning
problem as
\begin{subequations}
\label{eq:problem}
\begin{align}
\uSafe&(x) = \argmin_{u}\hspace{10pt}
 Q(x,u), \\
&\text{s.t.} \hspace{10pt}
 \gradBs(x)^{\Tr}[Ax + Bu]\le \gamma\Big(\frac{1}{\Bs(x)}\Big),	\label{eq:barrierConstraint} \\
&\hspace{25px}x(0) \in \IntC,
\end{align}
\end{subequations}
where $\Bs(x)$ is a user-defined candidate Lyapunov-like barrier function for the
constraint set $\C$ and $\gamma: \Re \rightarrow \Re$ is a class $\mathcal{K}$
function. Under the structure of value function in \eqref{eq:valueFunction}, we
observe that the optimization problem outlined in  \eqref{eq:problem} is convex
in the decision variable $u$. We formulate the Lagrangian function
$\mathcal{L}: \Re^{n} \times \Re^{m} \times \Re_{\ge 0}\rightarrow \Re$ as
\begin{equation}
\label{eq:1}
\begin{aligned}
\mathcal{L}(x,u,\nu) =&
Q(x,u)
+ \nu \Big[\gradBs(x)^{\Tr}[Ax + Bu] \\&
- \gamma\Big(\frac{1}{B(x)}\Big)\Big],
\end{aligned}
\end{equation}
where $\nu \in \Re_{\ge 0}$ is the Lagrange multiplier. The optimal control
for the constrained system can be obtained from
$\frac{\partial \mathcal{L}}{\partial u} = 0$, as
\begin{equation}
\label{eq:safeOptimal}
\uSafe(x) = -\Quu^{-1}\Qux x - \nu^{*}(x) R^{-1}B^{\Tr} \gradBs(x),
\end{equation}
where $\nu^{*}(x): \Re^{n} \rightarrow \Re_{\ge 0}$ is the Lagrange multiplier derived from the
Karush-Kuhn-Tucker(KKT) conditions \cite[Section 5.3.3]{boyd2004convex}, defined
as
\begin{equation}
  \label{eq:lambdaStar}
\nu^{*}(x) = \max\Big(\frac{C_{b}(x)}{R_{b}(x)},0\Big),
\end{equation}
where
$C_{b}(x) \triangleq
- \gradBs(x)^{\Tr}BR^{-1}B^{\Tr}\gradVStar(x)
+\gradBs(x)^{\Tr}Ax
+ \gamma(1/\Bs(x))$
(cf. \cite{almubarak2021CDC,tan2021TAC}), and
$R_{b}(x) \triangleq \quadr{BR^{-1}B^{\Tr}}{\gradBs(x)}$.
We observe that the expression for the optimal Lagrange multiplier
contains unknown terms of the system matrix $A$ and the matrix $P$ of
the optimal value function. To make the control law implementable and to
simplify the analysis, we estimate the Lagrange multiplier by a user-defined
positive constant $\nu = k_{sb} \in \Re_{>0}$. The certainty equivalence
controller thus becomes
\begin{equation}
\label{eq:finalControlLaw}
\uHat(x) = -\QuuHat^{-1}\QuxHat x - k_{sb} R^{-1}B^{\Tr} \gradBs(x),
\end{equation}
where $\QuxHat$ and $\QuuHat$ denote the online estimates for $\Qux$
and $\Quu$, respectively with appropriate dimensions.

\begin{rem}
The optimal Lagrange multiplier $\nu^{*}(\cdot)$ in \eqref{eq:safeOptimal} is a
state-varying gain that switches between zero and $C_{b}(x)/R_{b}(x)$ depending upon
the sign of $C_{b}(\cdot)$. If the
first term of the control input is sufficient to ensure safety
(i.e., satisfies \eqref{eq:barrierConstraint}) at a given state
$x \in \IntC$, then  $C_{b}(x) \le 0$ and consequently
$\nu^{*}(x) = 0$ (this property of Lagrange multipliers is termed as
complementary slackness, see \cite{boyd2004convex}). Additionally,
$\nu^{*}(\cdot)$ is non-zero when the first term of control input in
\eqref{eq:safeOptimal} is unable to satisfy the constraint on its own.
Thus, the second term also becomes active and
$\nu^{*}(\cdot)$ provides a way to ensure safety in a minimally-invasive
fashion. However, as discussed above, $\nu^{*}(\cdot)$ contains terms of
unknown/uncertain matrices $A$ and $P$. To make the controller implementable, we
approximate the multiplier $\nu^{*}(x)$ by a constant $\lamHat$. Under this
approximation, there is no way to switch-off the safety-inducing term
(second term of \eqref{eq:finalControlLaw}).
Thus, the proposed approximate control law
is only sub-optimal, with the optimality gap dependent on the choice of $\lamHat$.
Further, we show
that the satisfaction of the safety constraint is not
compromised under the approximation of the Lagrange multiplier by the constant
$k_{sb}$.
\end{rem}


\begin{rem}
The second term in \eqref{eq:finalControlLaw} closely resembles a ``safeguarding
controller'' coined in \cite{cohen2023Automatica}. The developments in the
present paper aim to bridge the gap between the ad-hoc approach of safeguarding
controllers and the theory of constrained optimal control.
\end{rem}

We now extend the actor-critic learning algorithm from \cite{vamvoudakis2017SysConLet} to learn the controller in
\eqref{eq:finalControlLaw} online.

\subsection{Actor-Critic based online learning }
The optimal $Q(x,u^{*})$ function from \eqref{eq:QStarDefinition} can be parameterized
as
\begin{equation}
\label{eq:Qparameterized}
Q(x,u^{*}) = \frac{1}{2} \quadr{\overline{Q}}{X} =
\underbrace{\frac{1}{2} \text{vech}(\overline{Q})^{\Tr}}_{\triangleq W_{c}^{\Tr}}\phi(X),
\end{equation}
where $\text{vech}(\overline{Q}) \in \Re^{p}$ with $p \triangleq (n+m)(n+m+1)/2$, denotes the half-vectorization of
$\overline{Q}$ yielding a column vector containing the upper-triangular
elements of $\overline{Q}$, where elements off the diagonal are considered to be
$2\overline{Q}_{ij}$; and
$\phi:\Re^{n+m} \rightarrow \Re^{p}$ denotes the quadratic
basis function yielding a vector containing the elements
$\{X_{i}X_{j}\}_{i \in \mathbb{N}_{n}, j \in \mathbb{N}_{m}}$.

Since the matrix $\overline{Q}$ is unknown, the vector $W_{c}$
and control law $u^{*}(\cdot)$ are unknown and unimplementable, and thus
require corresponding estimators. The ``critic'' estimator approximating the $Q$
function is given by
\begin{equation}
\label{eq:5}
\hat{Q}(x,\uHat) = \WcHat^{\Tr} \phi(X),
\end{equation}
where $\WcHat \in \Re^{p}$ is the weight estimate for the
critic and the estimated control law (``actor'') is given by
\begin{equation}
\label{eq:finalSafeControlLaw}
\uHat(x) = \WaHat^{\Tr} x - k_{sb} R^{-1} B^{\Tr} \gradBs(x),
\end{equation}
where $\WaHat \in \Re^{n \times m}$ is the actor weight estimate. The
objective of the actor and critic components of the algorithm is to minimize the estimation errors defined as
$\WaTilde(t) \triangleq - \Qxu \Quu^{-1} - \WaHat(t)$ and
$\WcTilde(t) \triangleq W_{c} - \WcHat(t)$  respectively.

To learn the ideal weights $W_{c}$ and $W_{a}(\triangleq -\Qxu\Quu^{-1})$, we write
the fixed point optimality equation based on integral reinforcement learning \cite{vamvoudakis2017SysConLet} as
\begin{equation}
\label{eq:fixedpoint}
\begin{aligned}
Q(x(t),u^{*}(t)) = &Q(x(t-T),u^{*}(t-T)) \\&
- \frac{1}{2} \int_{t-T}^{t}
(\quadr{M}{x} + u^{*\Tr}Ru^{*})
d\tau,
\end{aligned}
\end{equation}
where $T \in \Re_{>0}$ is a fixed time interval. The expression in \eqref{eq:fixedpoint} is
the continuous-time equivalent of the Bellman optimality equation for
discrete-time reinforcement learning. Now, under the uncertainties in the system
matrix $A$ and the ARE solution $P$, we define the temporal difference (TD) error in the estimate of the
$Q$ function as
\begin{equation}
\label{eq:9}
\begin{aligned}
e_{c}(t) &\triangleq \;
 \hat{Q}(x(t),\hat{u}(t))
-\hat{Q}(x(t-T),\hat{u}(t-T))  \\& \;\;
  +\frac{1}{2} \int_{t-T}^{t}
(\quadr{M}{x} + \quadr{R}{\hat{u}})
  d\tau \\&
= \WcHat^{\Tr} \psi(t)
  +\frac{1}{2} \int_{t-T}^{t}
(\quadr{M}{x} + \quadr{R}{\hat{u}})
  d\tau,
\end{aligned}
\end{equation}
where $\psi(t) \triangleq \phi(X(t)) - \phi(X(t-T))$.
To learn the ideal weights online, we define the squared norm of the
critic error as
$
\delta_{c} \triangleq \frac{1}{2} \norm{e_{c}}^{2},
$
and subsequently, write the gradient descent-based update law for the critic as
\begin{equation}
\label{eq:criticUpdate}
\begin{aligned}
\WcHatDot(t) &=
 -\eta_{c} \frac{\psi(t)}{(1 + \quadr{}{\psi(t)})^{2}} e_{c}(t),
\end{aligned}
\end{equation}
where $\eta_{c} \in \Re_{>0}$ is a user-defined gain. The update law for the
actor is given by
\begin{equation}
\label{eq:actorUpdate}
\begin{aligned}
\WaHatDot(t) &=
 \proj(-\eta_{a} (\QuxHat(t)^{\Tr} \QuuHat(t)^{-1} + \WaHat)),
\end{aligned}
\end{equation}
where the estimates $\QuuHat(t)$ and $\QuxHat(t)$ are extracted from
$\WcHat(t)$, $\proj(\cdot)$ denotes the projection operator \cite{lavretsky2013robust}  that ensures
$\norm{\WaHat(t)} \le \overline{W}_{a} \fall t \in \Re_{\ge 0}$ where
$\overline{W}_{a} \in \Re_{>0}$ is a user-defined bound and $\eta_{a} \in \Re_{>0}$
is the user-defined actor gain. Since the update law for actor depends on the
estimate of the critic, the critic's learning rate  must be substantially larger.

Under the update law in \eqref{eq:criticUpdate}, the time derivative of the
critic estimation error becomes
\begin{equation}
\label{eq:13}
\WcTildeDot(t) = -\eta_{c} \frac{\psi(t)\psi(t)^{\Tr}}{(1 + \quadr{}{\psi(t)})^{2}} \WcTilde(t).
\end{equation}
\begin{assm}
\label{assm:PE}
The signal $\frac{\psi(t)}{1+ \quadr{}{\psi(t)}}$ is persistently exciting
(PE).
\end{assm}

\subsection{Safety and Stability Analysis}

\begin{thm}
For the system in \eqref{eq:plant} and under the critic and actor update laws in
\eqref{eq:criticUpdate} and \eqref{eq:actorUpdate} respectively, the control law
in \eqref{eq:finalSafeControlLaw} ensures that the state $x$, the actor and
critic weight estimation errors ($\WaTilde$ and $\WcTilde$) are uniformly
ultimately bounded (UUB), and the set $\C$ is forward invariant.
\end{thm}

\begin{proof}
According to the definition of forward invariance
\cite{blanchini1999Automatica}, we initialize the state $x$ such that
$x(0) \in \IntC$.
We now consider the positive definite candidate Lyapunov function
$\mathcal{V}: \mathcal{D} \rightarrow \Re$, where
$ \mathcal{D} \triangleq \IntC \times \Re^{p+nm}$,
defined as
\begin{equation}
\label{eq:2}
\begin{aligned}
\mathcal{V}(\zeta) &=
\frac{1}{2}\quadr{P}{x}
+\lamHat B_{s}(x)
+\frac{1}{2}\norm{\WcTilde}^{2}
+\frac{1}{2} \trace(\quadr{}{\WaTilde}),
\end{aligned}
\end{equation}
where $\zeta \triangleq
[x^{\Tr} \;
\WcTilde^{\Tr} \;
\text{vec}(\WaTilde)^{\Tr}]^{\Tr}$ is the augmented state vector for the overall
closed loop system. Using \eqref{eq:boundsOfBarrier} one can show that there
exist two class $\mathcal{K}$ functions $\alpha_{l}$, $\alpha_{u}$ such that
$\alpha_{l}(\norm{\zeta}) \le \mathcal{V}(\zeta) \le \alpha_{u}(\norm{\zeta}) \fall \zeta \in \mathcal{D}$.
In other words, $\mathcal{V}(\zeta)$ is a valid candidate Lyapunov function
\cite{khalil2002nonlinear}.
The time derivative of the $\mathcal{V}(\cdot)$ under the control policy $\uHat(x)$ is
\begin{equation}
\begin{aligned}
\dot{\mathcal{V}}(\zeta) &=
x^{\Tr}P (Ax + B\uHat)
+\lamHat \gradBsT(Ax + B\uHat) \\&
-\WcTilde^{\Tr} \WcHatDot
-\trace(\WaTilde^{\Tr} \WaHatDot).
\end{aligned}
\end{equation}
Using \eqref{eq:are}, \eqref{eq:criticUpdate}, \eqref{eq:actorUpdate} and
substituting \eqref{eq:finalSafeControlLaw} we have
\begin{equation}
\begin{aligned}
  &\dot{\mathcal{V}}(\zeta)
  \le
    -\frac{1}{2} \quadr{(M+ \Qxu \Quu^{-1} \Qux + \Qxu \WaTilde^{\Tr})}{x}
  \\&
  - \eta_{c} \quadr{\frac{\psi \psi^{\Tr}}{(1 + \psi^{\Tr}\psi)^{2}}}{\WcTilde}
  -\lamHat^{2} \quadr{BR^{-1}B^{\Tr}}{\gradBs}
  \\&
  -\eta_{a}\trace(
\WaTilde^{\Tr} \WaTilde + \WaTilde^{\Tr}\QuxTilde^{\Tr} R^{-1} )
  +\lamHat \gradBsT A x
  \\&
  + \lamHat \gradBsT B \WaHat^{\Tr} x.
\end{aligned}
\end{equation}
Under Assumption \ref{assm:PE}, we can find the upper-bound
\begin{equation}
\begin{aligned}
  &\dot{\mathcal{V}}(\zeta)
  \le
    -\kappa_{x} \norm{x}^{2}
  -\kappa_{c} \norm{\WcTilde}^{2}
    -\kappa_{a} \norm{\WaTilde}^{2}
    \\&
  - \kappa_{b} \norm{B^{\Tr} \gradBs(x)}^{2}
  + \lamHat \upperB \norm{B^{\Tr} \gradBs(x)} + \iota_{c},
\end{aligned}
\end{equation}
where
$\kappa_{c} \triangleq \frac{1}{8}\eta_{c}$,
$\kappa_{a} \triangleq \eta_{a} $,
$\kappa_{x} \triangleq \frac{1}{2}
\eigMin(M+ \Qxu \Quu^{-1} \Qux)
$,
$\kappa_{b} \triangleq \lamHat^{2}\eigMin(R^{-1})$,
$\upperB \triangleq \sup_{x\in \IntC} \norm{Ax}/\norm{B} + \norm{\overline{W}_{a}x} $,
$\iota_{c}\triangleq \frac{2 \eta_{a}^{2}\overline{W}_{a}^{2}\norm{R^{-1}}^{2}}{\eta_{c}}
+ \sup_{x \in \IntC}\frac{\overline{W}_{a}^{2}\norm{x}^{2}}{2}
$
are positive constants. Completing the squares, we write
\begin{equation}
\begin{aligned}
  \dot{\mathcal{V}}(\zeta)
  &\le
    -\kappa_{x} \norm{x}^{2}
  -\kappa_{c} \norm{\WcTilde}^{2}
    -\kappa_{a} \norm{\WaTilde}^{2}
    \\&
  - \frac{\kappa_{b}}{2} \norm{B^{\Tr} \gradBs(x)}^{2}
  + \iota,
\end{aligned}
\end{equation}
where $\iota \triangleq \frac{\upperB^{2}}{2\lamHat \eigMin(R^{-1})} +\iota_{c} $ is a
finite positive constant. It can be shown that there exists a class
$\mathcal{K}$ function $\alpha_{v}(\cdot)$ such that
\begin{equation}
\begin{aligned}
 \alpha_{v}(\norm{\zeta})
  &\le
    \kappa_{x} \norm{x}^{2}
  +\kappa_{c} \norm{\WcTilde}^{2}
    +\kappa_{a} \norm{\WaTilde}^{2}
    \\&
  + \frac{\kappa_{b}}{2} \norm{B^{\Tr} \gradBs(x)}^{2}.
\end{aligned}
\end{equation}
We thus write
$
\dot{\mathcal{V}}(\zeta) \le - \alpha_{v}(\norm{\zeta}) + \iota
$. Now, since $x(0) \in \IntC$, $\mathcal{V}(\zeta(0))$ is a finite quantity.
Additionally, we observe that $\dot{\mathcal{V}}(\cdot) < 0$ outside the compact
set
$\Omega_{v} \triangleq \{\zeta \in \mathcal{D}: \norm{\zeta} \le \alpha_{v}^{-1}(\iota) \}$.
Thus using \cite[Theorem 4.18]{khalil2002nonlinear} it can be shown that
$\zeta$ is uniformly ultimately bounded (UUB). Since
$x(0) \in \IntC \implies \mathcal{V}(\cdot) < \infty \fall t \in \Re_{\ge 0}$,
the RCBF  $\Bs(x(t)) < \infty \fall t \in \Re_{\ge 0}$. By definition, at no
point in time does the state trajectory intersect the boundary of the safe set
$\dC$ \cite{tee2009Automatica}.  Thus the state
$x(t) \in \C \fall t \in \Re_{\ge 0}$ and the set $\C$ is forward invariant
for the system in \eqref{eq:plant}.
\end{proof}

\begin{rem}
The ultimate bound for the system depends on the term $\iota$ which can be
reduced by choosing the critic gain $\eta_{c}$ to be much larger than
$\eta_{a}$, and choosing an appropriate safety gain $\lamHat$. However, upon
increasing $\lamHat$, it can be shown that the control effort required increases. Thus
there exists a trade-off between the
safety and control effort objectives.
\end{rem}

\section{Simulation Results}
To demonstrate the safety and performance of the proposed control
algorithm, we consider the linear system with $A = [0 , 1; 1.6 , 2.8]$ and $B = [0;1]$.
We seek to solve the linear quadratic regulator problem with the matrices
$M = \mathbb{I}_{2}$ and $R = 0.1$. We impose a constraint on the norm of the
state $x$, i.e., $\norm{x(t)} \le 1.5 \fall t \in \Re_{\ge 0} $. To incorporate the constraint we construct the candidate RCBF as  $\Bs(x) = (\frac{1.5^2}{1.5^2 - \quadr{}{x}} - 1)^2$.
The actor
gain ($\eta_{a}$) was chosen to be 0.05, and the critic gain ($\eta_{c}$) was
chosen to be $20$. To enforce the safety constraints, the gain $\lamHat$ was
chosen to be $0.2$. The integration window $T$ was set to $0.01s$. We apply an
exploration noise to the control input for the first $10s$ of the simulation to
ensure sufficient excitation and enable state exploration.

\begin{figure}[htbp]
\centerline{\includegraphics[width=\linewidth]{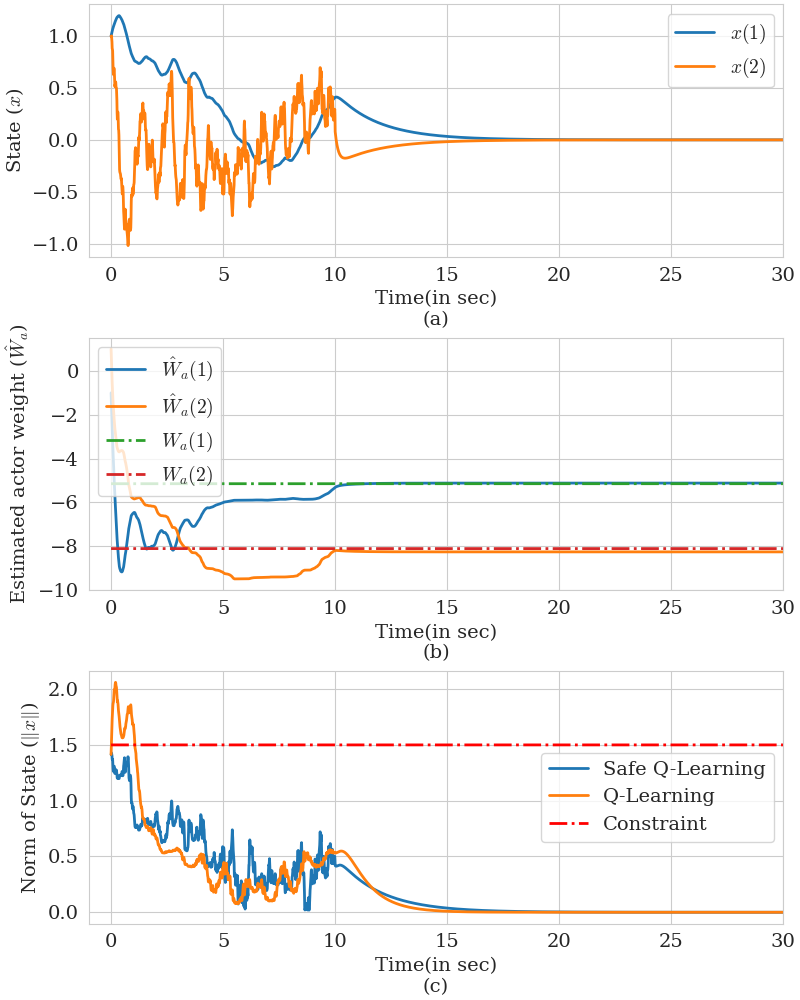}}
\caption[]{\label{fig:plot}
  (a) State Trajectory for the proposed algorithm
  (b) Estimated actor weights ($\WaHat$) compared with the true control gains ($W_{a}$)
  (c) Plot of the norm of state $x$ compared with algorithm from \cite{vamvoudakis2017SysConLet}.
}
\end{figure}

Fig. \ref{fig:plot}(a) shows the state trajectory of the system under the
influence of the proposed control law starting from the initial condition
$x_{0} = [1;1]$. We observe that the proposed controller
meets the regulation objectives and brings the state to the origin within
approximately 10 seconds after removing the exciting signal. Fig.
\ref{fig:plot}(b) shows the plot for the actor weight estimate ($\WaHat$). We
observe that the estimated actor weight converges close to the true control gain
($W_{a}$).
However, there is a small
steady-state error in the estimate $\hat{W}_{a2}$ (i.e., the weight
corresponding to $x_{2}$).
The plot for the norm of the state $x$ is shown in Fig.
\ref{fig:plot}(c) along with the plot corresponding to the controller in
\cite{vamvoudakis2017SysConLet}. We observe that during the initial phase of the
online
RL training, controller from \cite{vamvoudakis2017SysConLet} violates the safety
constraint, whereas the proposed controller meets the safety constraints at all
times.
\begin{table}[htbp]
\caption[]{\label{tab:cost} Performance analysis under different safety gains}
\centering
\resizebox{0.8\columnwidth}{!}{
\begin{tabular}[tb]{|c|c|c|}
  \hline
  Safety gain $\lamHat$ & Total cost & Peak control effort \\
  \hline
  \hline
  0.01 & 43.652 & 18.746 \\
  0.1 & 40.631 & 18.45 \\
  0.2 & 40.021 & 18.39 \\
  0.3 & 39.833 & 24 \\
  0.5 & 39.293 & 40 \\
\hline
\end{tabular}
}
\end{table}

\begin{figure}[htbp]
\centerline{\includegraphics[width=\linewidth]{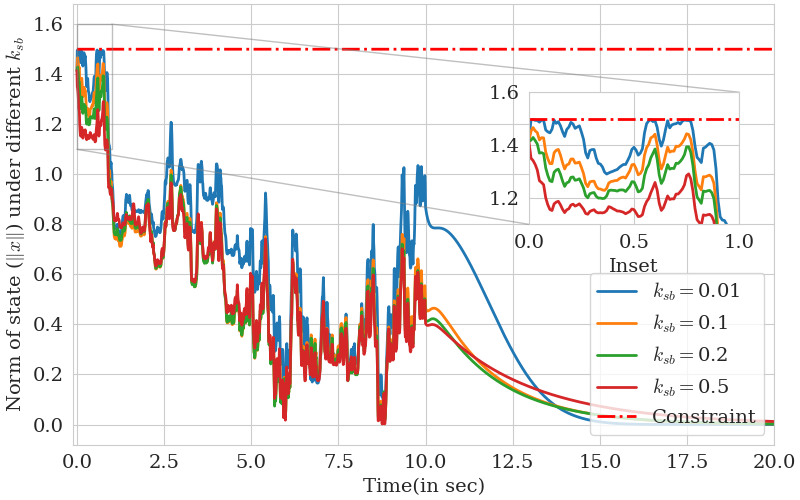}}
\caption[]{\label{fig:gainvar}
  The plot of the norm of the state $x$ for the proposed controller under different values of $\lamHat$.
}
\end{figure}

We now study the effect of variation of the safety gain $\lamHat$ on the
performance of the proposed algorithm. To demonstrate the same, we simulate the
linear system under different values of $\lamHat$ starting from the same initial
condition $x_{0} = [1;1]$. The exploration signal for all the cases was kept the same to enable a fair comparison. The plot for the norm of the
state, along with the constraint bound, is shown in Fig. \ref{fig:gainvar}. We
observe that the state ventures closer to the constraint boundary upon decreasing the value of $\lamHat$. In other words, upon increasing $\lamHat$, the
algorithm becomes more conservative.

The comparison of the cost and the peak control effort under different values
of $\lamHat$ is given in Table \ref{tab:cost}. We observe that upon increasing
$\lamHat$, the total cost decreases, but the peak control effort required
increases (although for lower values of $\lamHat$, the exploration control input dictates the peak control effort). Thus we observe a
trade-off between the cost and the control effort required. However, the safety constraint is never violated for all
values of $\lamHat$.

\section{Conclusions}
In this paper, we propose a safe Q-learning algorithm utilizing reciprocal
control barrier functions. Such an approach has the distinct advantage of
learning optimal control policies for uncertain LTI systems with user-defined
state constraints. We formulate the safe Q-learning problem as a constrained optimization problem involving a constraint
on the time derivative of the RCBF. Subsequently, we derive adaptation laws
based on integral reinforcement learning for
the actor and critic estimators to estimate the constrained optimal control law
online. We prove that under the proposed control law, the user-defined
constraint set is forward invariant. Additionally, the state and the estimation
errors are shown to be uniformly ultimately bounded via a Lyapunov analysis. We
subsequently demonstrate the safety and stability performance in a simulation study.
Future extensions to the present work could include
extending the safe Q-learning algorithm to consider the dynamics of the
Lagrange multiplier generated by the KKT conditions and
extending the proposed algorithm to include both actuation and state constraints.








\bibliography{uni.bib}
\bibliographystyle{ieeetr}

\end{document}